\documentstyle{article}

\newcommand{\nl}{\newline}

\title{Squeezed States \\
 and Helmholtz Spectra}

\author{Francisco Delgado C.
\thanks{Depto. de Matem\'aticas, ITESM, Campus Estado de M\'exico. \nl
A.P. 2, 52926, Atizap\'an, Edo. de Mex., M\'exico.}
, Bogdan Mielnik
\thanks{Institute of Theoretical Physics, Warsaw University. \nl
 ul. Hoza 69, 00-681, Warsaw, Poland.} \\ \\
and \\ \\
Marco A. Reyes.
\thanks{Instituto de F\'\i sica de la Universidad de Guanajuato, 
37150, Guanajuato, M\'exico.} \\ \\
Depto. de F\'\i sica, CINVESTAV, \\
A.P. 14-740, 07000, M\'exico D.F., M\'exico.}

\date {}

\begin{document}
\begin{titlepage}

\maketitle

\begin{abstract}

The 'classical interpretation' of the wave function $\psi (x)$ reveals an interesting operational aspect of the Helmholtz spectra. It is shown that the traditional Sturm-Liouville problem contains the simplest key to predict the squeezing effect for charged particle states.

\vskip 1cm

PACS number(s): 02.70.Hm, 41.75.-i, 03.65.Ge

\vskip 1cm

{\it Preprint} CINVESTAV: CIEA-FIS/97-07

\end{abstract}

\end{titlepage}

\section{Introduction.}
\setcounter{equation}{0}

One of the most transparent pictures of the
Scr\"odinger's eigenvalue equation in 1-dimension is
obtained by reinterpreting the space coordinate $x$ as the "time" ($x=t$),
and the values of the wave function $\psi (x)$ and its derivative $\psi '(x)$ as the
position and momentum of a classical point particle ~\cite{pru,mal,coleman,avron,bmar}. The
resulting classical trajectory obeys Hamilton equations which can be further simplified by introducing the Pr\"ufer's angle on the Poincare plane ~\cite{pru}.
The integration of the 1-st order "angular
equation" then turns out one of the most efficient methods to determine the energy
spectra ~\cite{pru, bmar}. In this note we show that the 'classical algorithm'
works even better for the Helmholtz problem ~\cite{lind,ghat, bender}; in addition, it
exhibits a new aspect of the Helmholtz spectra.

\section{The Helmholtz spectrum.}
\setcounter{equation}{0}

\hfill

Given an interval $[a,b] \subset {\bf R}$ and a real function
$\phi : [a,b] \rightarrow {\bf R}$, the {\it Helmholtz equation} is a particular case of a Sturm-Liouville problem (see ~\cite{ghat, bender}):

\begin{equation}
{d^2 \psi \over dx^2} + \lambda \phi (x) \psi (x) =0
\end{equation}

and the {\it Helmholtz spectrum} is defined  as the set of all $\lambda \in
{\bf R}$ for which (2.1) has a non trivial solution $\psi$ vanishing on both
ends:

\begin{equation}
\psi (a) = \psi (b) =0
\end{equation}

The 'trajectory picture' ~\cite{pru,bmar} applies quite naturally to the Helmholtz
eigenproblem, and the spectral condition is much simpler than
in the Schr\"odinger's case. Let $\psi(x)$ be any real solution of (2.1).
Define: $x=t$, $\psi=q$, $\psi ' =p$; eq. (2.1) then becomes:

\begin{eqnarray}
{dq \over dt} & = & p \nonumber \\
{dp \over dt} & = & - \lambda \phi (t) q
\end{eqnarray}

with the 'spectral condition':

\begin{equation}
q(a) = q(b) = 0
\end{equation}

Observe, that (2.3) are the canonical equations for a classical point-particle
with the Hamiltonian:

\begin{equation}
H(t) = {p^2 \over 2} + \lambda \phi (t) {q^2 \over 2}
\end{equation}

Each integral trajectory of (2.3) 'portraits' a wave function $\psi (x)$
and its first derivative. We shall further assume that $0 \not\equiv \phi (t) \ge0$ is a piece-wise continuous function
in $[a,b]$. The attractive oscillator force in (2.3) then drives the classical
trajectory clockwise around the origin of the phase plane (Fig. 1). When $\lambda$ grows to
$+ \infty $, the end point $(q(b), p(b))$ must cross the $p-$ axis an infinite
number of times (Fig. 2). Each time this happens, $\lambda$ belongs to the point
spectrum of (2.1). In this way the Helmholtz eigenproblem (2.1) has a discrete spectrum
in the form of an infinite ladder of positive eigenvalues determined by the classical trajectories (compare the theorems ~\cite{hoch, contqm}).

The calculation of the spectral values can be carried out by 
introducing the polar variables on the phase plane: $p= \rho \cos{\alpha}$,
$q = \rho \sin{\alpha}$. The variable $\alpha$ then separates and (2.3) reads
~\cite{pru}:

\begin{eqnarray}
\dot \alpha & = & \cos^2{\alpha} + \lambda \phi(t) \sin^2{\alpha}  \\
{\dot \rho \over \rho} & = & {1 - \lambda \phi{(t)} \over 2} \sin{2 \alpha}
\end{eqnarray}

with the initial and spectral conditions:

\begin{eqnarray}
{\it initial \quad condition} & \alpha (0)=0 & \quad \\
{\it spectral \quad condition} & \Delta \alpha = \alpha (b) = n \pi & (n=0, \pm 1, \pm 2, \dots)
\end{eqnarray}

Some exact solutions of (2.1-2) were recently studied by Goyal et al to check the validity of the WKB and Langer approximations ~\cite{ghat}. For curiosity, we applied the angular method to the same collection of cases.
Eq. (2.6) was integrated numerically for a net of $\lambda$-values
by applying the Runge-Kutta method
and the condition (2.9) selected the eigenvalues $\lambda$ quoted in Table 1.

\begin{center}
\begin{tabular}{|l|l|l|l|l|l|}
\hline
$\phi(t)$   & $[a, b]$ &  \multicolumn{3}{c|}{Eigenvalues $\lambda$}  & Amplification    \\
\cline{3-5}
               &      &  'Exact'       &  Approximate  &  Angular     & constant $\sigma$  \\   \hline
$(t+\pi)^4$& $[0, \pi]$ &  0.00174401  &  0.00188559 (WKB)  &  0.00174401  &  -0.543046  \\ 
               & & 0.00734865  &   0.00754235 (WKB)    &  0.00734843  &   0.51828   \\  
               & & 0.0167524   &   0.0169703 (WKB)     &  0.0167517   &  -0.510087  \\  
\hline


$1.1 e^t -1$   & $[0,1]$ &  11.145735 & 11.145736 (LAN)  &  11.145735   &  -0.591942  \\  
               & & 47.301484   &   47.301493 (LAN)   &  43.301493   &   0.54752   \\  
               & & 108.428802   &  108.428616 (LAN)  & 108.428629   &  -0.52727   \\ 
\hline
$(1+\sin 2 \pi t)^2$   & $[0, 1]$    &    &  &5.3347146 &  -2.35707   \\ &    &     &    & 34.1068933 &  3.1455  \\
 &     &     &    &  86.8947093 & -3.42553  \\ \hline 
\end{tabular}
\end{center}
{\bf Table 1:} Helmholtz eigenvalues known in the literature compared with the results of the angular method. In addition we have examined $\phi (t)=\lambda (1+\sin{2 \pi t})^2$ in $[0,1]$. The last column gives the '{\it optical information}', disregarded in the literature.

\hfill

As one can see, for the low eigenvalues the ranking is won by the almost
ignored angular algorithm (both, in number of digits and simplicity of operation). Its additional advantage is that
it uncovers the links of the Helmholtz spectrum with the techniques of generating the squeezed states.

\section{The trajectory optics.}
\setcounter{equation}{0}

\hfill

In a typical Sturm-Liouville problem, one looks for a solution of a differential equation which fulfills a definite \it boundary condition \rm . The second independent solution is dimissed as unphysical. However, we shall show that for (2.1) the second solution too carries an important  message. As eqs. (2.3) are linear, the corresponding $q(t)$, $p(t)$
depend linearly on the initial values $q(a)$, $p(a)$:

\begin{equation}
\left(\matrix{q(t) \cr p(t) \cr}\right) = u(t,a) \left( \matrix{ q(a) \cr p(a) \cr} \right)
\end{equation}

where $u(t,a)$ is the simplectic {\it evolution matrix}
~\cite{gil, bmar} (also: the {\it transfer matrix}). The existence of
a non-trivial solution of (2.2) with $q(a)=q(b)=0$ means that the evolution matrix $u(a,b)$ in
the interval $[a,b]$ maps:

\begin{equation}
\left(\matrix{0 \cr p(a) \cr}\right)  \quad \quad \rightarrow \quad \quad \left( \matrix{ 0 \cr p(b) \cr} \right)
\end{equation}

and since $u(a,b)$ is simplectic, it must be of the form:

\begin{equation}
u(b,a) = \left( \matrix{ \sigma & 0 \cr \eta & 1/\sigma \cr} \right) ; \quad \quad \sigma, \eta \in  {\bf R}, \sigma \ne 0
\end{equation}

Once this is known, one can deduce not only the spectral orbit of (2.3), but also the form of other
trajectories. Consider therefore the congruence of 
trajectories which diverge at $t=a$ from the initial point
($q_0, p$), where $q_0$ is fixed and $p$ varies. If $\lambda$ is an
eigenvalue of (2.1) and $u(b,a)$ is of the form (3.3), then at $t=b$:

\begin{equation}
\left( \matrix{q(b) \cr p(b) \cr} \right) =
\left( \matrix{\sigma & 0 \cr \eta & 1/\sigma \cr} \right)
\left( \matrix{q_0 \cr p \cr} \right) =
\left( \matrix{\sigma q_0 \cr p' \cr} \right)
\end{equation}

i.e., all orbits departing from the same $q_0$ (with different momenta p) meet at
$\sigma q_0$ (with momenta $p'=\sigma q_0 + p/ \sigma$). Thus, the eigenvalues $\lambda$ of (2.1) are the real numbers for which the time evolution (2.3) focuses any congruence of classical
orbits diverging from any common initial $q_0$. The optical constants $\sigma$, $\eta$ can be read from the second (disregarded) solution of (2.1).

Note that the focusing,
in general, is not the \it scale transformation \rm ~\cite{brown}, (as the matrix (3.3) is
triangular), though it is an example of squeezing ~\cite{yuen, marho, manko}. The scale
transformation ($\eta=0$) is an exceptional effect which can happen only for some $\phi$'s (see ~\cite{ours}), whereas
the {\it focusing}  is a typical phenomenon which must occur for any positive $\phi(x)$,
on an infinite ladder of $\lambda$'s.
The converse problem is as well interesting. Instead of fixing an interval and looking for $\lambda$, fix any $\lambda >0$ and $\beta (t) \ge 0$ and consider an expanding time interval $[0,\tau]$. If $\phi (t)$ mantains some positive average in $[0, \infty)$, the final point $(q(\tau ), p( \tau))$, as $\tau \rightarrow \infty$ must cross infinity of times the $p$ -axis, each time generating our focusing effect no matter the exact form of $\beta (t)$. Should $\beta (t)$ be a random variable performing errating oscillations, the motions (2.3) will be chaotic too; yet, for any congruence of trajectories, a clean 'echo effect' of the initial configuration (an amplified or reduced image)   must emerge now and then against the chaotic forces. Note also that the effect must have a quantum equivalent.
 
\section{The Quantum Images.}
\setcounter{equation}{0}

\hfill

Consider the Hamiltonian $H(t)$ in form (2.5), where $q$ and
$p$ are position and momentum operators. Since (2.5)
is cuadratic, the quantum evolution follows faithfully the classical motions. The techniques of the classical "piloting trajectory" gives specially good results for the Gaussian wave packets ~\cite{manko, ge}. We shall show, that it provides also an explicit expression for the evolution operator $U(t,a)$:

\begin{equation}
{dU(t,a) \over dt} = -i H(t) U(t,a); \quad U(a,a)=1
\end{equation}

With this end define:

\begin{equation}
W(t)=e^{i \alpha (t) H_0} U(t,a)
\end{equation}

where $H_0={1 \over 2}(p^2+q^2)$ is the traditional oscillator Hamiltonian and $\alpha (t)$ is a real, differentiable function. Due to the well known transformation rules: $e^{i \alpha H_0} q e^{-i \alpha H_0} = q \cos \alpha + p \sin \alpha$, $e^{i \alpha H_0} p e^{-i \alpha H_0} = p \cos \alpha - q \sin \alpha$, the operator $W(t)$ fulfills:

\begin{eqnarray}
{dW \over dt}& = & i [(\dot \alpha - \cos^2 \alpha - \lambda \phi (t) \sin^2 \alpha) {p^2 \over 2} + \nonumber \\
& & + (\dot \alpha - \sin^2 \alpha - \lambda \phi (t) \cos^2 \alpha) {q^2 \over 2} + \\
& & + (1-\lambda \phi (t)) \sin \alpha \cos \alpha {qp +pq \over 2}] W(t) \nonumber
\end{eqnarray}

Notice, that the time dependent term coupled to $p^2/2$ corresponds exactly to the Pr\"ufer equation. Thus, if  $\alpha (t)$ is the Pr\"ufer angle  (2.6), then (4.3) reduces to:

\begin{eqnarray}
{dW \over dt}& = & i [{1- \phi(t) \over 2} \sin 2 \alpha {qp+pq \over 2} + \nonumber \\
& & + (1-\lambda \phi (t)) \cos 2 \alpha {q^2 \over 2}] W(t)
\end{eqnarray}

and integrates explicitely in terms of the Pr\"ufer's radious $\rho$:

\begin{equation}
W(t) = e^{i \ln ({\rho (t) \over \rho (a)})  {qp+pq \over 2}} e^{i \eta (t) {q^2 \over 2} }
\end{equation}

Inserting (4.5) into (4.4) and remembering that the first exponent in (4.4) is the scale operation, one gets:

\begin{equation}
\eta (t) = \int_a^t (1-\lambda \phi (t)) ({\rho (a) \over \rho (t)})^2 \cos 2 \alpha dt
\end{equation}

and returning to (4.2) one has an explicit formula for the unitary operator $U(t,a)$ with three exponential terms piloted by Pr\"ufer variables (2.6-7):

\begin{equation}
U(t,a) = e^{-i \alpha (t) H_0} e^{i \ln ({\rho (t) \over \rho (a)}) {qp +pq \over 2}} e^{i \eta (t){q^2 \over 2}}
\end{equation}

If in addition the number $\lambda$ is the eigenvalue of the Sturm-Liouville problem (2.1) $\Rightarrow \alpha (b) = n \pi$ ($n=1,2,...$), then the first factor in (4.7) becomes $P^n$ (where $P=parity$) and

\begin{equation}
U(b,a) = P^n e^{i \ln ({\rho (b) \over \rho (a)}) {qp +pq \over 2}} e^{i \eta {q^2 \over 2}}
\end{equation}

transforms any initial wave packet $\psi (x) = \psi (x,a)$ into:

\begin{equation}
\psi_U (x) = [U(b,a) \psi] (x) = {1 \over \sqrt{\sigma}} e^{i \eta {x^2 \over 2 \sigma^2}} \psi (x / \sigma)
\end{equation}

where $\sigma=\pm \rho (a) / \rho (b)$. The probability density $\rho=|\psi (x)|^2$ is simultaneously mapped into its own diffractionless image:

\begin{equation}
\rho_U = |U \psi|^2 = {1 \over |\sigma|} \rho (x/ \sigma)
\end{equation}

It thus turns out that the authors looking for the squeezing parameters were indeed solving the XIX-century spectral problem ! The relative simplicity of the three term formula (4.7) (compare  ~\cite{gil, brown, yuen, marho}) is due to the "bottom structure" (2.6-9). Its additional 
implication concerns the "potential kicks" $\delta (t) V(q)$ considered recently as a method of controlling and cooling quantum systems ~\cite{kick}. While the "kicks" are mathematically simple, they are overidealized ( no $\delta$-pulses available in laboratories!). Notice, however, that if $\lambda$ is an eigenvalue of (2.1) and if $\phi (t)$ is symmetric $\phi (-t) = \phi (t)$ in a symmetric operation interval $[-a,a]$, then the eigenvector $\psi$ (and consistenly the phase trajectory $q(t))$ is even/odd in $[-a,a]$, implying the trivial squeezing $\sigma=\pm 1$, and the only non trivial part of (4.7-8) is $P^n e^{i \eta {q^2 \over 2}}$. For $n=2 l$, this is the unitary operator describing the results of a sharp kick of the oscillator potential $-\eta \delta (t) q^2 /2$. Thus, the spectrum  of (2.1) tells in addition how to "imitate softly" the effects of $\delta$-kicks converting the results on kicked oscillators ~\cite{kick} into a branch of realistic technology.

\vskip 0.5cm

\section{Magnetic Images.}
\setcounter{equation}{0}

\vskip 0.5cm

The  results (3.4), (4.3) have a simple laboratory model.
Consider a non-relativistic charged particle in 3-dim, in a homogeneous
time-dependent magnetic field ${\bf B} (t) = B(t) {\bf n}$, with a vector
potential:

\begin{equation}
{\bf A} ({\bf x}, t) = - {1 \over 2} {\bf x} \times {\bf B} (t)
\end{equation}

where {\bf n} is a constant unit vector and $B(t)= B \gamma(t/T)$ is the magnetic field intensity, $B$ defining 'the amplitude' and  $\gamma$ the
'shape' of the magnetic pulses.  If {\bf n}
point the $z$-axis, the Schr\"odinger's wave
packet obeys the Hamiltonian:

\begin{eqnarray}
H(\tau) & = & {1 \over 2} p_3^2 - \beta \gamma(\tau) M_3 +  \\
        &   & + {1 \over 2} [p_1^2 + p_2^2 + \beta^2 \gamma^2(\tau) (q_1^2+q_2^2)]  \nonumber
\end{eqnarray}

which we have written in the dimensionless coordinates:
$\tau=t/T$, $(q_1,q_2,q_3)= \sqrt{m/T \hbar} (x,y,z)$, $(p_1,p_2,p_3)=(\sqrt{m \hbar /T})^{-1} (p_x,p_y,p_z)$;
all physical information now contained in the 'shape function' $\gamma (\tau)$
and in the 'dimensionless amplitude':

\begin{equation}
\beta = {e B T \over 2 m c}
\end{equation}

In (5.2) the term  ${1 \over 2} p_3^2$ means the free propagation along the $z$-axis,
the $- \beta \gamma(\tau) M_3$ generates rotations in the $q_1, q_2$-plane. The last term (a 2-dimensional oscillator),
splits into two twin 1-dimensional oscillators of the form:

\begin{equation}
h(\tau)=p^2 / 2 + \beta^2 \gamma^2(\tau) q^2 / 2
\end{equation}

where $(q,p)$ means either $(q_1,p_1)$ or $(q_2,p_2)$.
Quite evidently, (5.4) is a model of (2.5) with
$\lambda=\beta^2$ and $\phi(\tau)=\gamma^2(\tau) \ge 0$. If $\gamma(\tau) \not\equiv 0$,
the spectrum of the Helmholtz equation (2.1) forms a ladder of eigenvalues, and
if $\lambda = \beta^2$ is one of them, both oscillators yield the simultaneous focus of the trajectory
congruencies (in classical case) or probability patterns (in quantum case).
To check this, we have asked the computer to draw congruencies of classical
trajectories diverging from a common initial point, with a common
initial $v_z$, for $\lambda=\beta^2$ coinciding with one of the eigenvalues of (2.1) and
with $\phi(\tau)= \beta^2 (\tau)$ taken from Table 1. In all cases we have obtained quite
precise optical images on planes perpendicular to the solenoid axis (see Fig. 3). The dependence of the phenomenon on the Sturm-Liouville spectrum is illustrated by sequences of amplified focuses corresponding exactly to the nodal points of the wave function (2.1-2). Henceforth, the integration of the first order differential equation of Pr\"ufer (for one variable) offers the simplest algorithm to predict the {\it optical echo} (squeezing) for charged particle beams.

The "reduced images"  might imply new 
schemes to produce narrow particle beams. The "amplfied images" are no less interesting. They can certainly be used to separate the wave packet components in the quantum state engineering ~\cite{ours, wolf, dav, poy}). The magnetic pulses $B(t)$ with $\phi (t)=\gamma^2 (t)$
 symmetric in the operation interval and with $\lambda = \beta^2$ in the spectrum of (2.1), provide a magnetic model for systems under the influence of oscillator kicks.

{\bf Acknowledgements.} The support of CONACYT, M\'exico, is acknowledged.

\newpage

\centerline{\bf Figure Captions}
\vskip 1cm

{\bf Fig. 1.-} The orbit of the classical oscilator (2.5) for $\phi (t) = \sin^2 t $, $\lambda = 10$. The points of intersection with the $p$-axis determine the time moments $t_1=1.43610, \quad t_2=2.53311, \quad t_3=4.46714, \quad t_4=5.51549, ...$, and the amplifications $\sigma_1=-0.52322, \quad \sigma_2=0.65528, \quad \sigma_3=-0.23823, \quad \sigma_4=0.27464, ...$ of the optical images. The similar phenomenon would be generated by a chaotic $\phi {(t)}$.

\vskip 1cm

{\bf Fig. 2.-} The orbits of (2.3) and the Helmholtz spectrum of (2.1). If the end point of
the orbit with q(a)=0 places itself on the $p$-axis, $\lambda$ is an
eigenvalue of the Helmholtz equation (1). Above: a sequence of orbits
reaching $q(b)=0$ after multiple intersections with the $p$-axis illustrates the sequence of
Helmholtz eigenstates for $\phi(t)= (1+\sin 2 \pi t)^2$,  $[a,b]= [0, 1]$. The corresponding optical constants are reported in Table 1.

\vskip 1cm

{\bf Fig. 3.-} The optical effect of the Helmholtz spectrum observed in the solenoid for sinusoidally oscilating $\gamma (\tau) = 1+\sin 2 \pi \tau$. {\it a)} When $\lambda= \beta ^2=5.3347146$ is
an eigenvalue of (2.1), the congruence of particle trajectories forms an exact optical focus at $\tau=1$ ($t=T$) (in quantum case, an amplified diffractionless image) with $\sigma = -2.35707$. The integration of the first order differential equation of Pr\"ufer for one angular variable turns out the simplest method to predict the effect. {\it
b)} For $\lambda=\beta^2=86.8947093$ the congruence forms $3$ increasingly amplified focuses (images of the initial point) in exact correspondence to the $3$ nodal points of the wave function $\psi (x)$.

\end{document}